\documentclass[%
 reprint,
superscriptaddress,
nofootinbib,
 amsmath,amssymb,
 aps,
]{revtex4-1}

\usepackage{mhchem}

\usepackage[utf8x]{inputenc} 
\usepackage[T1]{fontenc} 
\usepackage{float}

\usepackage[symbol]{footmisc}

\usepackage{graphicx}
\usepackage{dcolumn}
\usepackage{bm}
\usepackage{hyperref}

\begin{document}


\title{Water-induced formation of an alkali-ion dimer in cryptomelane nanorods}



\author{Shaobo Cheng}
\thanks{These authors contributed equally to this work.}
\affiliation{%
Department of Condensed Matter Physics and Materials Science, Brookhaven National Laboratory, Upton, NY 11973, USA.
}%

\author{Vidushi Sharma}
\thanks{These authors contributed equally to this work.}
\affiliation{%
 Department of Physics and Astronomy, Stony Brook University, Stony Brook, NY 11794-3800, USA.
}%
\affiliation{%
 Institute for Advanced Computational Science, Stony Brook University, Stony Brook, NY 11794-5250, USA
}%

\author{Altug S. Poyraz}
\affiliation{%
Energy Sciences Directorate, Brookhaven National Laboratory, Upton, NY 11973, USA.
}%
\affiliation{%
Department of Chemistry and Biochemistry, Kennesaw State University, Kennesaw, GA 30144, USA.
}%

\author{Lijun Wu}
\affiliation{%
Department of Condensed Matter Physics and Materials Science, Brookhaven National Laboratory, Upton, NY 11973, USA.
}%

\author{Xing Li}
\affiliation{%
Department of Chemistry, Stony Brook University, Stony Brook, NY 11794, USA.
}%

\author{Amy C. Marschilok}
\author{Esther S. Takeuchi}
\affiliation{%
Energy Sciences Directorate, Brookhaven National Laboratory, Upton, NY 11973, USA.
}%
\affiliation{%
Department of Chemistry, Stony Brook University, Stony Brook, NY 11794, USA.
}%
\affiliation{%
Department of Materials Science and Chemical Engineering, Stony Brook University, Stony Brook, NY 11794, USA.
}%
\author{Kenneth J. Takeuchi}
\affiliation{%
Department of Chemistry, Stony Brook University, Stony Brook, NY 11794, USA.
}%
\affiliation{%
Department of Materials Science and Chemical Engineering, Stony Brook University, Stony Brook, NY 11794, USA.
}%

\author{Marivi Fern\'{a}ndez-Serra}
\email{maria.fernandez-serra@stonybrook.edu vidushi.sharma@stonybrook.edu}
\affiliation{%
 Department of Physics and Astronomy, Stony Brook University, Stony Brook, NY 11794-3800, USA.
}%
\affiliation{%
 Institute for Advanced Computational Science, Stony Brook University, Stony Brook, NY 11794-5250, USA
}%

\author{Yimei Zhu}
\affiliation{%
Department of Condensed Matter Physics and Materials Science, Brookhaven National Laboratory, Upton, NY 11973, USA.
}%


\date{\today}

\begin{abstract}
Tunneled metal oxides such as $\alpha-$Mn$_8$O$_{16}$ (hollandite) have proven to be compelling candidates for charge-storage materials in high-density batteries. In particular, the tunnels can support one-dimensional chains of K$^+$ ions (which act as structure-stabilizing dopants) and H$_2$O molecules, as these chains are favored by strong H-bonds and electrostatic interactions. In this work, we examine the role of water molecules in 
enhancing the stability of K$^+$-doped $\alpha-$Mn$_8$O$_{16}$ (cryptomelane). The combined experimental and theoretical analyses show that for high enough concentrations of water and tunnel-ions, H$_2$O displaces K$^+$ ions from their natural binding sites. This displacement becomes energetically favorable due to the formation of K$_2^+$ dimers, thereby modifying the stoichiometric charge of the system. These findings have potentially significant technological implications for the consideration of cryptomelane as a Li$^+$/Na$^+$ battery electrode. Our work establishes the functional role of water in altering the energetics and structural properties of cryptomelane, an observation that has frequently been overlooked in previous studies.
\end{abstract}

\maketitle

\section{Introduction}
The highly earth abundant and environmentally friendly manganese oxide compounds with a tunneled structure have drawn much attention in wide ranging applications in the realm of multiferroic materials, catalysts, super-capacitors, and batteries \cite{ref1, ref2, ref3, ref4, ref5, ref6}. 
Owing to the large $[2 \times 2]$ tunnel sizes, $\alpha-$Mn$_8$O$_{16}$, which possesses a hollandite structure, could accommodate and reversibly insert/retract large species such as Na$^+$, K$^+$, Ag$^+$ and H$_2$O \cite{ref7, ref8, ref9, ref10, ref11, COCKAYNE201253, ref23}. 
As shown in Fig. \ref{fig:fig1}a and b, potassium doped $\alpha-$Mn$_8$O$_{16}$ (cryptomelane) contains a framework of edge and corner shared MnO$_6$ octahedra \cite{ref12, ref13}. 
The relatively large ionic radius of K$^+$ ions and the strong electrostatic repulsion between the adjacent K$^+$ ions prevent full occupation of all the tunnel sites \cite{ref14}.
Different from Ag$^+$ ions with relatively small ionic radii occupying the 2a Wyckoff position, K$^+$ ions can occupy the 2b Wyckoff position, because the 2b position has larger space (as shown in Fig. \ref{fig:fig1}c) \cite{ref15, ref16}. 

To date, the question of how the filling of the tunnels might be modified by the presence of water has received very little attention, even despite the fact that water plays an important role in a wide spectrum of materials science studies \cite{ref17, ref18}. In battery chemistry, water molecules can act as a shielding component to facilitate ion diffusion and reduce polarization.
Although the lattice positions of H$_2$O within cryptomelane have not been analyzed in detail in previous studies, the presence of H$_2$O has been inferred to affect the overall electrochemistry of the material \cite{ref19}. Previous research suggests that water can enhance the structural stability of host structures during reversible ion intercalation/deintercalation processes \cite{ref20}.
Water can be removed from the sample by exposure to high temperatures, but exposure to aqueous solutions or even ambient air will result in the readsorption of water molecules \cite{ref21, ref22}.
A strong electrostatic attraction exists between the O-lone pair of electrons of H$_2$O and the K$^+$ ions, making incorporation of H$_2$O in the $[2 \times 2]$ tunnel favorable.
Despite the obvious physical reasons to quantify the effects that water might have on the performance of these materials, the difficulty of producing, measuring and comparing samples with and without water has hindered further research in this direction. 

In this work we investigate the impact of incorporating water and K$^+$ in $\alpha-$Mn$_8$O$_{16}$ using complementary approaches based on density functional theory (DFT) simulations and transmission electron microscopy (TEM) experiments.
The experiments show evidence of the displacement of K$^+$ ions from their natural 2b Wyckoff position to a 2a site upon hydration. Our simulations show that this displacement
can occur only when there are enough H$_2$O molecules in the channel, which compete with K$^+$ ions to occupy the larger 2b sites. As a consequence the K$^+$ ions not only move to the less favorable 2a sites, but do so by forming a di-cation molecule, which has a larger double ionization energy than the ionization energy of two independent K atoms. 

The paper is organized as follows. We first briefly summarize the simulation and experimental methods used in the study. Then we present the experimental results, which evidence the K$^+$ ions structural sites in samples with and without water content.
The experiments are followed by a theoretical section where  we analyze the interplay between K$^+$ ions and H$_2$O, and show that the incorporation of H$_2$O into the tunnels induces a large displacement of the K$^+$ ions thereby influencing their kinetics and mobility inside the tunnels. The DFT simulations provide an in-depth analysis of how different doping levels affect the structure and thermodynamics of the channels. 
Finally, a combined experimental and theoretical analysis section explaining each other's findings is presented.

\section{Methods}
Structural \textit{ab initio} DFT calculations were performed using the Spanish Initiative for Electronic Simulations with Thousands of Atoms (SIESTA) code \cite{ref26}. The core electrons were described by norm-conserving pseudopotentials generated by the Troullier–Martins approach \cite{ref27}. A numerical atomic orbitals (NAO) basis set with double-$\zeta$ polarization was used for the valence electrons. In this work, we used the gradient-corrected exchange-correlation (XC) functional, vdW-BH,  as implemented in SIESTA to include the effects of van der Waals interactions in the system \cite{ref28}. The unit cell lattice parameters for all the K- and H$_2$O- doped hollandite structures were chosen to be $a = b = 9.81$ \AA\, and $c = 2.90$ \AA. The supercells were constructed by stacking 2 to 5 unit cells of $\alpha-$Mn$_8$O$_{16}$ along the $c$-axis to obtain various concentrations of dopants (K$^+$ and H$_2$O) in the tunnel. The convergence of structural geometries and forces was tested for a $2 \times 2 \times 10$ Monkhorst-Pack \textit{k} grid and an energy cutoff of 350 Ry. The structures were fully optimized such that both the atomic positions and lattice vectors were relaxed until the remnant forces in the system were less than 0.04 eV \AA$^{-1}$. 

The transmission electron microscopy work was conducted at Brookhaven National Laboratory with a JEOL ARM-200 CF machine. The HAADF images were acquired with the collection angle from 67 to 275 mrad. 
The cryptomelane nanorods were synthesized using hydrothermal methods and the chemical component is identified by inductive coupled plasma (ICP) as K$_{0.8}$Mn$_8$O$_{16}\bm{\cdot} y$H$_2$O \cite{ref24}. The dry samples were further annealed at $200 ^{\circ}$C under $\sim 10^{-4}$ Torr vacuum for 1 day in the sealed furnace. 

\section{Results and discussion}
\subsection{Experimental results}
\begin{figure*}
    \centering
    \includegraphics[width=5.in]{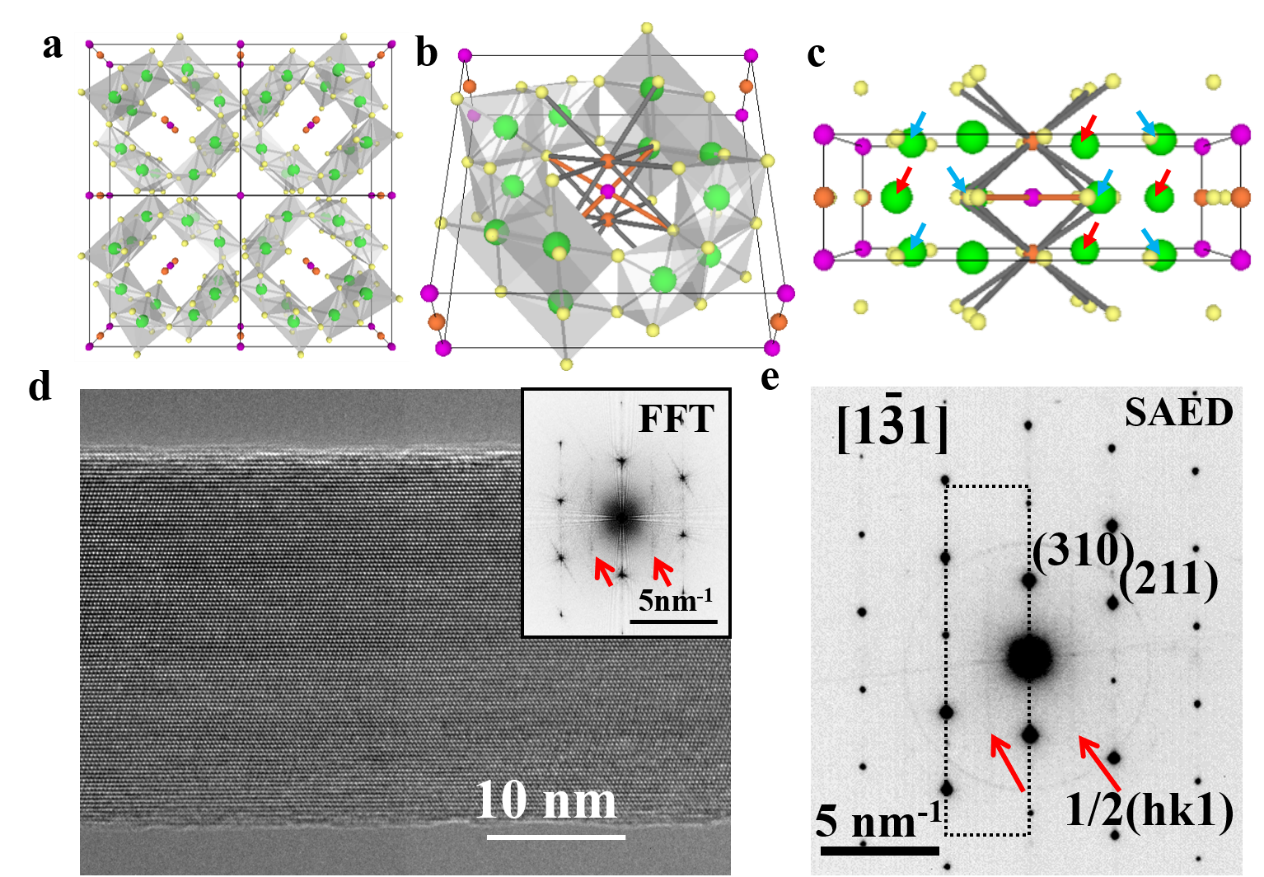}
    \caption{Structure and microscopy observations of the cryptomelane nanorod. Perspective view of the cryptomelane atomic models from the (a) $[001]$, (b) $[104]$, and (c) $[100]$ projection. Orange, purple, green, and yellow spheres represent K$^+$ at 2b, K$^+$ at 2a, Mn, and O, respectively. The thin black rectangle outlines the unit cell. Thick grey and orange lines in (b) and (c) outline the K$^+$(2b)-O$^{2-}$ and K$^+$(2a)-O$^{2-}$ bond, respectively. K$^+$(2b) bonds with 8 surrounding O$^{2-}$ with a bond length of 2.91 \AA, while K$^+$(2a) forms bond with 4 surrounding O$^{2-}$ with bonding length of 2.56 \AA. The red and blue arrows highlight Mn atoms adjacent to K$^+$(2b) and K$^+$(2a), respectively. (d) Typical high-resolution TEM image of a cryptomelane nanorod from $[1\overline{3}1]$ zone axis. The corresponding FFT result is also shown in the inset. (e) Selected area electron diffraction pattern acquired from the same nanorod, which shares the similar streaking feature with the FFT result in (d) due to the 2c ordering. Streaks at $1/2$c are indicated by red arrows.}
    \label{fig:fig1}
\end{figure*}

A high resolution TEM image of a hydrated cryptomelane nanorod is presented in Fig. \ref{fig:fig1}d. Individual rods share similar morphologies with an average diameter of 20-50 nm and an average length of several micrometers. The nanorod is single-crystalline with its long axis along the $[001]$ direction. The selected area electron diffraction (SAED) pattern for this nanorod is shown in Fig. \ref{fig:fig1}e. The extra streaks in the SAED pattern, as marked by the red arrows, along the $[001]^*$ direction can be observed and located at the center of $hk\emph{0}$ row and $hk\emph{1}$ (or $hk\bar{\emph{1}}$) row, indicating a 2$c$ ($c$: $c$-lattice parameter) ordering along the $[001]$ direction. Given the presence of water, water and K$^+$ ions should be alternatively arranged along the $c$ direction, forming 2$c$ ordering. The elongation of the streaks along the $hk\emph{0}$ direction indicates that the coherent length of the ordered structure is short along in-plane (or $ab$) directions. The streaks can also be observed in the Fast Fourier Transform (FFT) result as shown in the upper-right corner inset of Fig. \ref{fig:fig1}d, consistent with the SAED result.

\begin{figure*}
    \centering
    \includegraphics[width=3.8in]{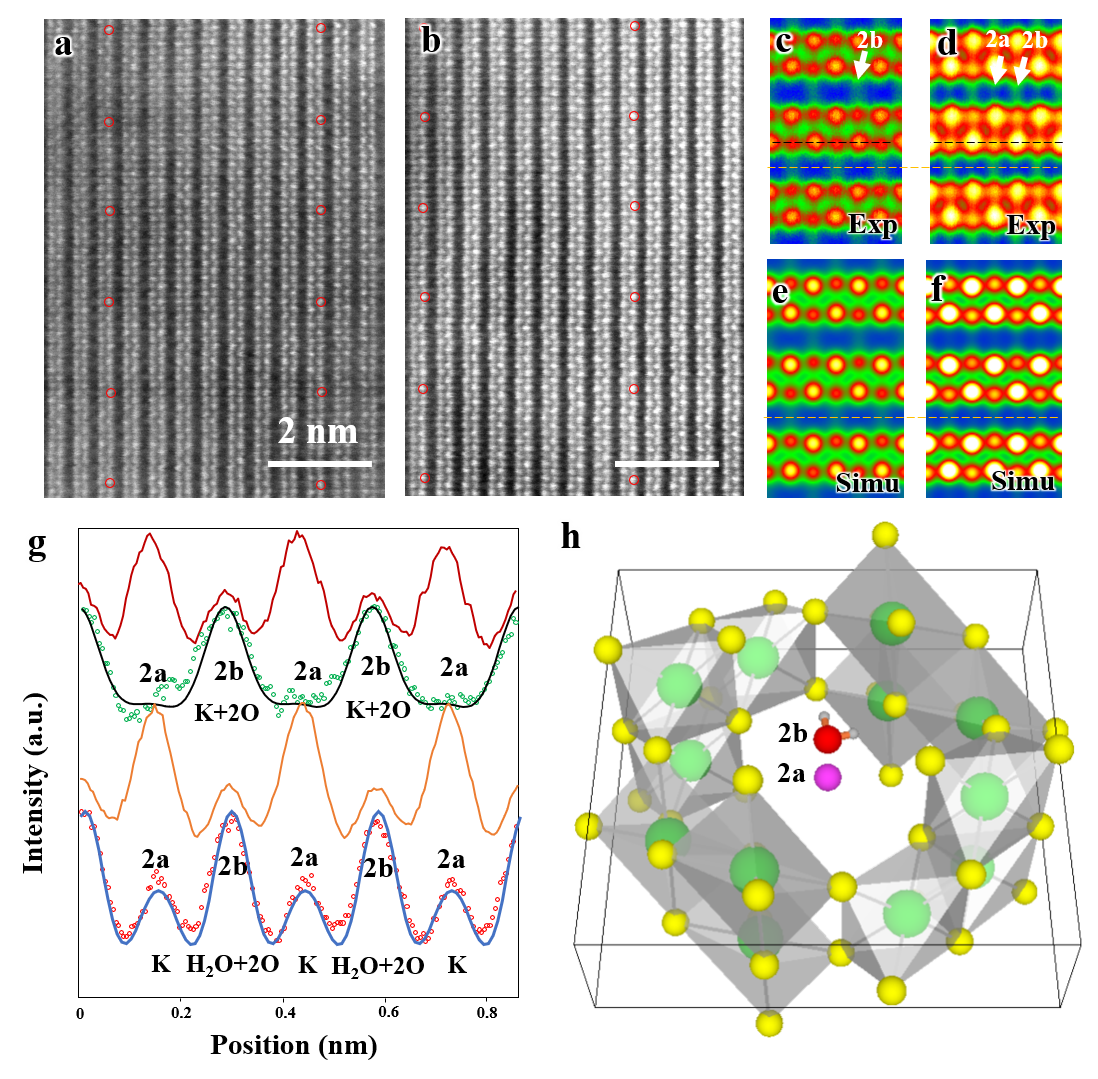}
    \caption{Comparison of dry and hydrated cryptomelane nanorods from in-plane direction. (a) Typical HAADF image of a cryptomelane nanorod. (b) HAADF image for hydrated cryptomelane. Both images were recorded along $[100]$ direction. (c), (d) Averaged images over 5 units in (a) and (b), as marked by red circles in (a) and (b). The averaged images are rotated 90$^\circ$ with respect to the original STEM images. The images are shown in false color with intensity values indicated by the blue-green-red-yellow-white color map in an ascending order. (e), (f) Simulated images of the structural model with thicknesses of (e) 32 nm and (f) 42 nm. In (e), 0.4 K$^+$ occupy 2b sites. In (f), 0.7 H$_2$O occupy 2b sites, 0.2 K$^+$ occupy 2b sites and 0.2 K$^+$ occupy 2a sites, respectively. (g) Intensity profiles of horizontal scan lines shown in (c)-(f). The dark-red and orange lines are from Mn column (black horizontal lines in (c), (d)), while the green and red dots are from the tunnels (orange lines in (c), (d)). The black and blue lines are the calculated intensities of (e) and (f). The intensity is normalized for clarity. (h) An atomic model showing water located at the 2b site, with K$^+$ at the 2a site.}
    \label{fig:fig2}
\end{figure*}

The typical high angle annular dark field scanning transmission electron microscopy (HAADF-STEM) images of dry and hydrated cryptomelane nanorods are compared in Fig. \ref{fig:fig2}a,b, respectively. To improve the signal-to-noise ratio, the images are averaged over several small units, as shown in Fig. \ref{fig:fig2}a,b, indicated by red circles. Fig. \ref{fig:fig2}c,d are the corresponding averaged images from Fig. \ref{fig:fig2}a,b with 90$^{\circ}$ rotation, whose intensity values are indicated by the blue-green-red-yellow-white color map in the ascending order. One typical feature of an HAADF-STEM image is that the image contrast is proportional to $Z^{1.7}$, where $Z$ is the effective atomic number \cite{ref3,ref25}. Mn has the largest atomic number ($Z_{\text{Mn}}$=25) in the crystal, and thus exhibits the strongest contrast. Projected along the $[100]$ direction, the Mn adjacent to K$^+$ (2a) appears to have a higher contrast than the one adjacent to K$^+$ (2b) due to it overlapping with 3 surrounding O atoms as shown in Fig. \ref{fig:fig1}c. In Fig. \ref{fig:fig2}a,c, there is no contrast at the 2a site (adjacent to the strong contrast Mn), but a weak contrast at the 2b site (adjacent to the weak contrast Mn). Fig. \ref{fig:fig2}g shows the intensity profiles of the Mn column (red) and tunnel (green dots), respectively, with no obvious peak at the 2a site. These indicate that K$^+$ ions in dry samples only occupy the 2b sites in this region. For further verification, we performed HAADF-STEM image simulations based on the multi-slice method with the frozen phonon approximation. A K$_{0.8}$Mn$_8$O$_{16}$ atomic model with K$^+$ (occupancy=0.4) at 2b was used for the HAADF-STEM image simulations. The simulated image and intensity profile are presented in Fig. \ref{fig:fig2}e and the black line in Fig. \ref{fig:fig2}g, respectively, consistent with the experimental results. It is worth mentioning that a ``dry sample’’ does not mean that it is entirely devoid of water. However, a small amount of water does not affect the observations here.

Fig. \ref{fig:fig2}b shows the HAADF-STEM image for hydrated cryptomelane taken from the [100] direction. Similarly, Fig. \ref{fig:fig2}d shows the corresponding averaged images of Fig. \ref{fig:fig2}b with a 90$^{\circ}$ rotation. Apart from the contrast at the 2b site, the weak contrast also shows up at the 2a site, appearing as the small peaks at the 2a site in the intensity profile (red dots) in Fig. \ref{fig:fig2}g. We notice that there are two oxygens overlapping at the 2b site for pristine $\alpha-$Mn$_8$O$_{16}$, while there is no oxygen at the 2a site in the $[100]$ projection (Fig. \ref{fig:fig1}c). If the 2a site is occupied by H$_2$O, the enhanced contrast will be weak, due to the low $Z$ value of H$_2$O. Therefore, the contrast at 2a sites indicates that relatively heavier atoms occupy the 2a sites. A series of image simulations suggest that there will be a noticeable contrast enhancement at the 2a site provided that only a certain amount of K$^+$ ions occupy the 2a sites. The electron energy-loss spectroscopy (EELS) results acquired from both dry and hydrated samples are shown in Fig. S2, supplemental material. The prepeaks of O K edge in manganites reflect the concentration of oxygen \cite{PhysRevB.93.054409}. In the hydrated sample, the prepeak is higher, indicating more water in the hydrated nanorods.

\begin{figure*}
    \centering
    \includegraphics[width=5.in]{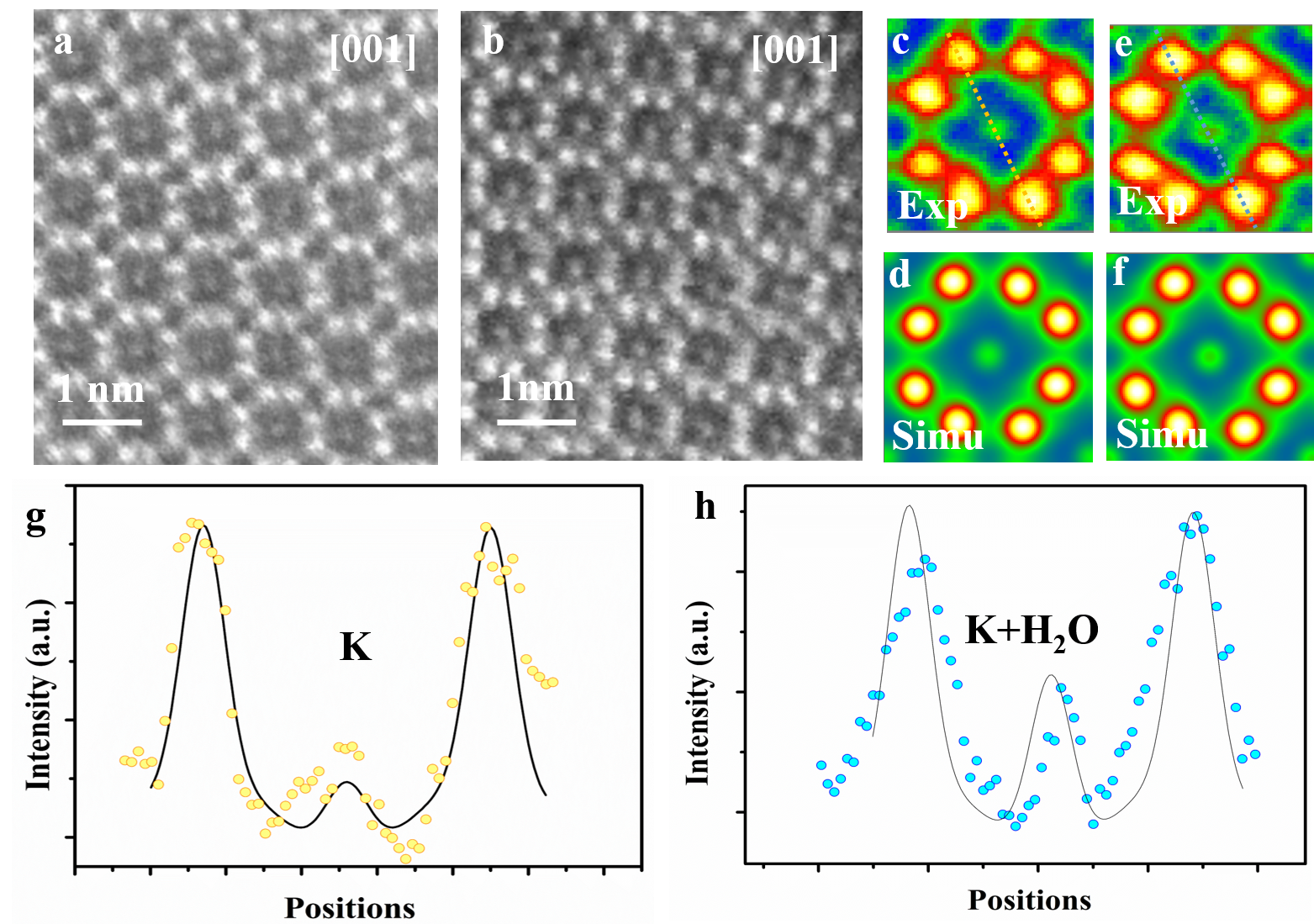}
    \caption{High resolution HAADF-STEM images for hydrated and dry potassium cryptomelane nanorods viewed along $[001]$ zone axis. (a) High resolution experimental image of a typical microtome-prepared dry cryptomelane sample. All the tunnels are well-aligned and occupied. (b) The high-resolution image from the hydrated sample. (c) Averaged image from (a). (d) Simulated HAADF-STEM image for K$_{0.80}$Mn$_8$O$_{16}$ unit cell with 80 nm thickness. (e) Averaged image from (b). (f) Simulated image for K$_{0.80}$Mn$_8$O$_{16} \bm{\cdot} (1.40)$H$_2$O with 80 nm thickness along $[001]$ direction. (g) Yellow dots and solid black line are intensity profiles for experimental and simulation results at the same position (i.e. dashed yellow line area in (c)), respectively. (h) Intensity profiles for the same area (i.e. blue dashed line area in (e)). The simulated relative intensity matches well with the experimental results.}
    \label{fig:fig3}
\end{figure*}

The microtome method was adopted for preparing the samples to view along the $[001]$ zone axis. The thickness of the microtome-prepared samples is set to be 80 nm and the corresponding HAADF-STEM image simulations are then carried out. The cross-section view of a typical dry cryptomelane nanorod is shown in Fig. \ref{fig:fig3}a. The basic cryptomelane structure is clearly resolved, where K$^+$ and eight surrounding Mn columns can be clearly observed. As opposed to Ag-doped hollandite, the intensity of K$^+$ columns is almost unchanged, indicating that every tunnel is approximately uniform in occupancy \cite{ref3}. As shown in Fig. \ref{fig:fig3}b for hydrated sample, the intensity of K$^+$ columns is overall brighter than that in Fig. \ref{fig:fig3}a. This tunnel intensity enhancement is due to the higher occupation of water, since the doping levels of K$^+$ ions are the same for both hydrated and dry samples. Fig. \ref{fig:fig3}c and e are averaged images over Fig. \ref{fig:fig3}a and b, respectively. To quantify the occupancy of water, a series of HAADF-STEM images simulations were carried out based on the structural model with different amounts of water. Fig. \ref{fig:fig3}d and f are the resulting simulated images of K$_{0.80}$Mn$_8$O$_{16}$ and K$_{0.80}$Mn$_8$O$_{16} \bm{\cdot} (1.40)$H$_2$O, respectively. Quantitative comparisons of the experimental and simulated atomic intensity profiles of K$_{0.80}$Mn$_8$O$_{16} \bm{\cdot} (1.40)$H$_2$O and K$_{0.80}$Mn$_8$O$_{16}$ are plotted in Fig. \ref{fig:fig3}g and h. The yellow and blue dots in Fig. \ref{fig:fig3}g and h are the intensity profiles from the yellow and blue line areas in Fig. \ref{fig:fig3}c and e. The black lines are the simulated intensity profiles from the same areas in Fig. \ref{fig:fig3}d and f, which agree well with experimental findings. It is worth mentioning that the background intensities of HAADF-STEM images are nonzero. Thus the simulated images are calibrated and then compared with the experimental ones.

\subsection{Theoretical calculations}
The experimental findings can be explained by DFT-based studies, which crucially also furnish new insights into the physics of these doped structures. The Wyckoff position occupied by K$^+$ inside the $[2 \times 2]$ $\alpha-$Mn$_8$O$_{16}$ tunnels is identified, and the binding energy of K$^+$ at its equilibrium position is computed for varying concentrations of H$_2$O. Additionally, the binding energy of H$_2$O at its optimal position in the large $\alpha-$Mn$_8$O$_{16}$ tunnel is found with and without K$^+$ ions inside the tunnel. 
We specify the K$^+$ and H$_2$O concentrations in terms of fractions per unit cell: the notation K$_{x}$Mn$_8$O$_{16} \bm{\cdot} y$H$_2$O signifies $x$ amounts of K$^+$ and $y$ amounts of H$_2$O per unit cell of hollandite. The cells are extended along the $c$-axis (from 2 to 5 unit cells) to attain various dopant concentrations. 
The framework of $\alpha-$Mn$_8$O$_{16}$ supports two $[2 \times 2]$ tunnels per unit cell, but all DFT calculations are performed for structures containing K$^+$ and H$_2$O in a single tunnel. Note that experiments consider the occupancy of the neighboring $[2 \times 2]$ tunnel as well, and moreover, a given sample shows dopants (K$^+$ and H$_2$O) present in both the tunnels. In our theoretical calculations however, we always keep one tunnel empty.
We have confirmed that dopants in one tunnel do not influence the energetics or geometries of those present in the other tunnel (see Fig. S3, supplemental material). We are therefore justified in inferring all possible energetics by combining results from multiple concentrations within a single tunnel. The number of dopants is varied ranging from a dilute limit, where the K$^+$ concentration ($x$) in K$_{x}$Mn$_8$O$_{16} \bm{\cdot} y$H$_2$O is $x<0.50$ to a concentrated limit in which $x>0.50$ in the hydrated tunnel. The concentration of water ($y$) is increased up to $y=1.00$ in the dilute-K regime and up to $y=0.60$ in the concentrated-K regime, in suitable steps of $x$ and $y$.

In cryptomelane (K$_{x}$Mn$_8$O$_{16}$), K$^+$ ions prefer to adopt a ‘2b’ Wyckoff position inside the hollandite tunnel ($x$ = 0.50, as shown in Fig. S4, supplemental material). This is arguably the most energetically favorable position for K$^+$ because it coordinates with a maximum number of oxygen (VIII) from the surrounding $\alpha-$Mn$_8$O$_{16}$ framework. The binding energy of K$^+$ in K$_{0.50}$Mn$_8$O$_{16}$ is found to be $-4.528$ eV. Similarly, in the hollandite structure containing water (Mn$_8$O$_{16} \bm{\cdot}y$H$_2$O), H$_2$O favors the tunnel-centered ‘2b’ Wyckoff site ($y$ = 0.50, as shown in Fig. S5a, supplemental material) but is weakly bound compared to K$^+$. The binding energy of H$_2$O in Mn$_8$O$_{16} \bm{\cdot}(0.50)$H$_2$O is $-0.212$ eV. The large difference between the binding energies of K$^+$ and H$_2$O is attributed to the characteristically different forces that bind the two species to the hollandite structure. While K is ionized inside the tunnel (the binding energy of K is very close to its ionization potential) and is electrostatically bound to the O’s belonging to the hollandite framework, the higher (less negative) binding energy of H$_2$O reflects the formation of non-bonded interactions of hydrogen bond (H-bond) type. Upon increasing the concentration of H$_2$O in the tunnel, an H-bond chain (Fig. S5b, supplemental material), spiraling down the $c$-axis of the tunnel is formed, resulting in an increased binding (BE$_{\text{H$_2$O}} = -0.475$ eV for Mn$_8$O$_{16} \bm{\cdot}(0.75)$H$_2$O). Therefore, H-bond cooperativity in water together with the one-dimensional $\alpha-$Mn$_8$O$_{16}$ tunnel-structure largely favors the addition of water molecules to maximize their coordination.

On incorporating H$_2$O and K$^+$ in the same tunnel, the structural energetics and optimal dopant positions differ in the dilute and concentrated regimes. In the dilute limit ($x$, $y < 0.50$ in K$_{x}$Mn$_8$O$_{16} \bm{\cdot}y$H$_2$O), both H$_2$O and K$^+$ occupy ‘2b’ Wyckoff positions ($x$, $y = 0.50$, in Fig. S5c, supplemental material) and the resultant structure is more stable, consistent with the 2$c$ ordering seen in Fig. \ref{fig:fig1}d,e. Water creates a solvation shell around K$^+$ and the lone-pair of electrons (from H$_2$O) proximate to K$^+$ ions further improves the coordination environment of the optimally-positioned K$^+$. Additionally, water binds to K$^+$ and lowers its binding energy. These dopant positions are quite robust but depend on their relative concentrations. K$^+$ prefers to maintain its more favorable ‘2b’ position at lower concentrations of water. The energy profile inside a K$_{0.50}$Mn$_8$O$_{16}\bm{\cdot}(0.50)$H$_2$O tunnel (Fig. S6, supplemental material) is consistent with this observation. In the presence of multiple K$^+$ ions and H$_2$O molecules in the tunnel, water is preferentially trapped between the recurring K$^+$’s, minimizing the electrostatic repulsion between them. This, in turn, enables us to incorporate more K$^+$ into the structure and approach the concentrated regime without compromising the energetics of the system.

\begin{figure}
    \centering
    \includegraphics[width=3.5in]{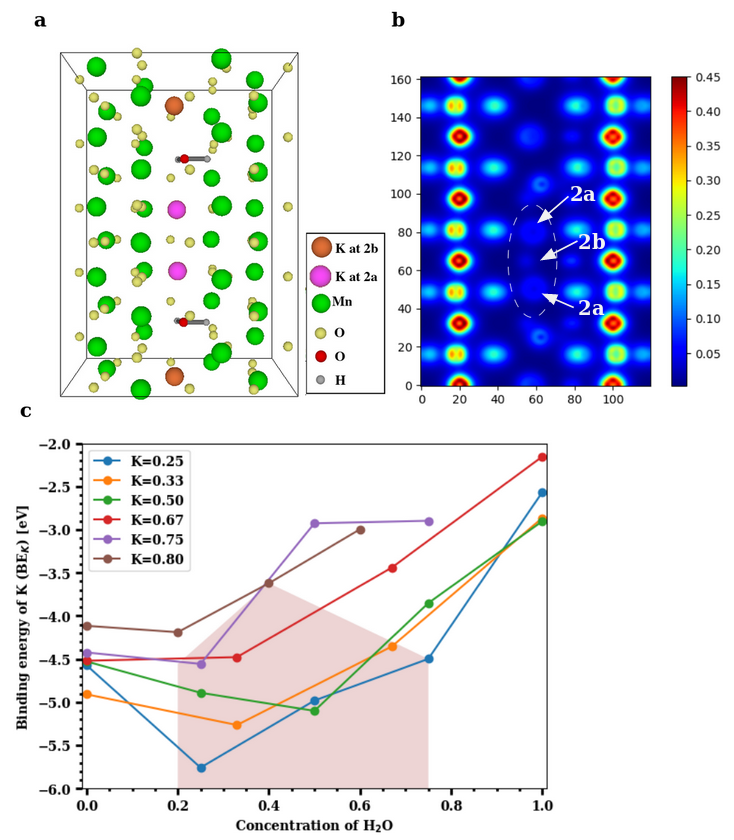}
    \caption{Results from DFT simulations of hydrated cryptomelane. (a) K$_{0.80}$Mn$_8$O$_{16}\bm{\cdot}(0.40)$H$_2$O: four K$^+$ and two H$_2$O in five unit cells of $\alpha-$Mn$_8$O$_{16}$ in a single tunnel, shown along the \textit{a}-axis. The two outer K$^+$’s occupy ‘2b’ sites; the two inner K$^+$’s are displaced by H$_2$O’s to ‘2a’ sites. One H$_2$O solvates two K$^+$’s – one above and one below. (b) Mean charge density plot for K$_{0.80}$Mn$_8$O$_{16}\bm{\cdot}(0.40)$H$_2$O from the $[100]$ axis, showing the displacement of K$^+$’s from ‘2b’ to ‘2a’ sites by H$_2$O’s inside the  tunnel. The tunnel-centered sites corresponding to red-colored charge densities are ‘2b’ sites, while those corresponding to yellow densities are ‘2a’ sites. (c) Binding energy of K$^+$ at varying concentrations of H$_2$O for K$_{x}$Mn$_8$O$_{16}\bm{\cdot}y$H$_2$O. The blue, orange and green curves indicate low-$K^+$ concentrations; red, purple and brown indicate high-K$^+$ concentrations in the tunnel. Experimentally-realized concentrations are closest to the brown curve.}
    \label{fig:fig4}
\end{figure}

In the high-concentration regime, our DFT based geometry optimization reveals that H$_2$O displaces K$^+$ from a ‘2b’ to a ‘2a’ Wyckoff position, in agreement with experimental results. The structure in Fig. \ref{fig:fig4}a, that is, K$_{0.80}$Mn$_8$O$_{16}\bm{\cdot}(0.40)$H$_2$O, displays a concentration closest to what is experimentally realized. The two H$_2$O molecules not only mimic each other’s orientation in the tunnel, but also play a very similar role. In particular, they solvate the K$^+$’s above and below, and push the inner central K$^+$’s toward the ‘2a’ site that was originally unfavorable at lower dopant concentrations. We therefore propose a necessary condition for water to displace K$^+$ from its originally favored position, namely that the combined fractional concentration of the dopants in the tunnel should exceed 1. Equivalently, the total number of K$^+$ and H$_2$O combined inside a  hollandite tunnel should exceed the available ‘2b’ sites in the tunnel.  In this setting, it is energetically viable for K$^+$ to be shifted to a ‘2a’ site, where the lack of O-coordination from the surrounding Mn$_8$O$_{16}$ framework is compensated by water. Water, on the other hand, prefers to remain at its ‘2b’ position, frequently forming an H-bond with the neighboring O of the enclosing Mn$_8$O$_{16}$ framework. In all our calculations the Mn$_8$O$_{16}$ framework was allowed to relax, and the screening of K$^+$ is achieved by a combination of both water (dominant) and small but necessary structural changes of the atoms within the channel.

The binding energy profile for K$_{x}$Mn$_8$O$_{16}\bm{\cdot}y$H$_2$O is plotted in Fig. \ref{fig:fig4}c. For low K$^+$ concentrations ($x \leq 0.50$) inside the $\alpha-$Mn$_8$O$_{16}$ tunnel, the minimum value of the binding energy is attained for equal concentrations of H$_2$O and K$^+$ inside the tunnel. By contrast, for higher K$^+$ concentrations ($x > 0.50$), the binding energy minimum is attained at lower H$_2$O concentrations in the tunnel. This is due to the excess of dopants in the tunnel. K$^+$ has a relatively larger ionic radius (2.80 \AA) and enriching the tunnel with K$^+$ and H$_2$O leads to an overall increase in energy due to steric effects. However, it is found that in K-rich hollandite, water -- in low amounts -- stabilizes the structure relative to structures containing no water. Additionally, as is shown in Fig. S8, supplemental material, H$_2$O does not facilitate the transport of K$^+$’s inside the $\alpha-$Mn$_8$O$_{16}$ tunnels, and overall, tends to stabilize K$^+$’s in the originally-chosen favorable coordination environment. Fig. \ref{fig:fig4} shows results computed via DFT, for the case of dopants occupying only one $[2 \times 2]$ hollandite tunnel. Results of double-channel occupancy for configurations comparable with experiments, are presented in Table \ref{tbl:1}. These are systems in which the dopant concentration in one of the tunnels is restricted to [K$^+$]=0.80 and [H$_2$O]=0.40. The neighboring tunnel contains K$^+$ and H$_2$O  with concentrations denoted by the region in Fig. \ref{fig:fig4}c, shaded in brown color. In particular, K$_{0.80}$Mn$_8$O$_{16}\bm{\cdot}(1.40)$H$_2$O in Table \ref{tbl:1} corresponds to [K$^+$]=0.80 and [H$_2$O]=0.40 in the first tunnel, whereas in the second tunnel, one has [H$_2$O]=1 and no K$^+$.

\begin{table}[b]
\caption{\label{tbl:1}Binding energies of K for various combined concentrations of K and H$_2$O occupying the two tunnels of $\alpha-$Mn$_8$O$_{16}$. [K] and [H$_2$O] indicate the total K$^+$ and H$_2$O concentrations summed over the two tunnels of hollandite.}
\begin{ruledtabular}
\begin{tabular}{ccc}
  [K] & [H$_2$O] & [BE$_{K}$ (eV)]  \\ 
  \colrule
  0.80 & 1.40 & -3.621 \\ 
  1.05 & 0.65 & -4.689 \\ 
  1.05 & 0.90 & -4.301 \\ 
  1.05 & 1.15 & -4.059 \\ 
  1.13 & 0.73 & -4.443 \\ 
  1.13 & 1.07 & -3.987 \\ 
  1.30 & 0.65 & -4.256 \\ 
  1.30 & 0.90 & -4.361 \\ 
  1.47 & 0.73 & -4.049 \\ 
  1.55 & 0.65 & -4.089  \\
\end{tabular}
\end{ruledtabular}
\end{table}

\subsection{Analysis and physical origin of the ion displacement}
Both experimental and simulation results indicate that K ions are displaced to ‘2a’ sites for high enough water content. However, this displacement is difficult to explain considering the strong Coulomb repulsion that the ions will experience in this location.
In order to establish the main driving force behind the displacement of K to ‘2a’ sites, we analyze the charge distribution in a channel with the displaced K configuration. Two different pairs of K atoms may be identified, namely a translated K$_a$$\cdots$K$_a$ pair occupying ‘2a’ sites, and a K$_b$$\cdots$K$_b$ pair occupying the originally favored ‘2b’ sites. K atoms are ionized (oxidized) within the tunnel, becoming K$^+$. The lost electron is delocalized over the neighboring O’s from the surrounding hollandite framework. However, when two K’s are stabilized in such close proximity as it happens in our system -- where $d_a=2.62$ \AA\, and $d_b=2.98$ \AA\, -- their interactions can give rise to the homonuclear ion K$_2^+$. This would imply that the actual charge in the cryptomelane tunnels is less than the nominal amount of K’s in the tunnel. This suggested formation of K$_2^+$, an elusive and highly unstable species, is facilitated by the presence of water in the tunnels. The H$_2$O’s provide not only a sufficient amount of electrostatic screening between the K$\cdots$K pairs, but also make the formation of K$_2^+$ energetically viable. One might expect a similar outcome when the tunnels are doped with Li$^+$/Na$^+$, which will deteriorate the performance of $\alpha-$Mn$_8$O$_{16}$. These large cations dimerize, and consequently, the effective charge is lower than the stoichiometric charge of the system \cite{ref10}.

\begin{figure*}
    \centering
    \includegraphics[width=5in]{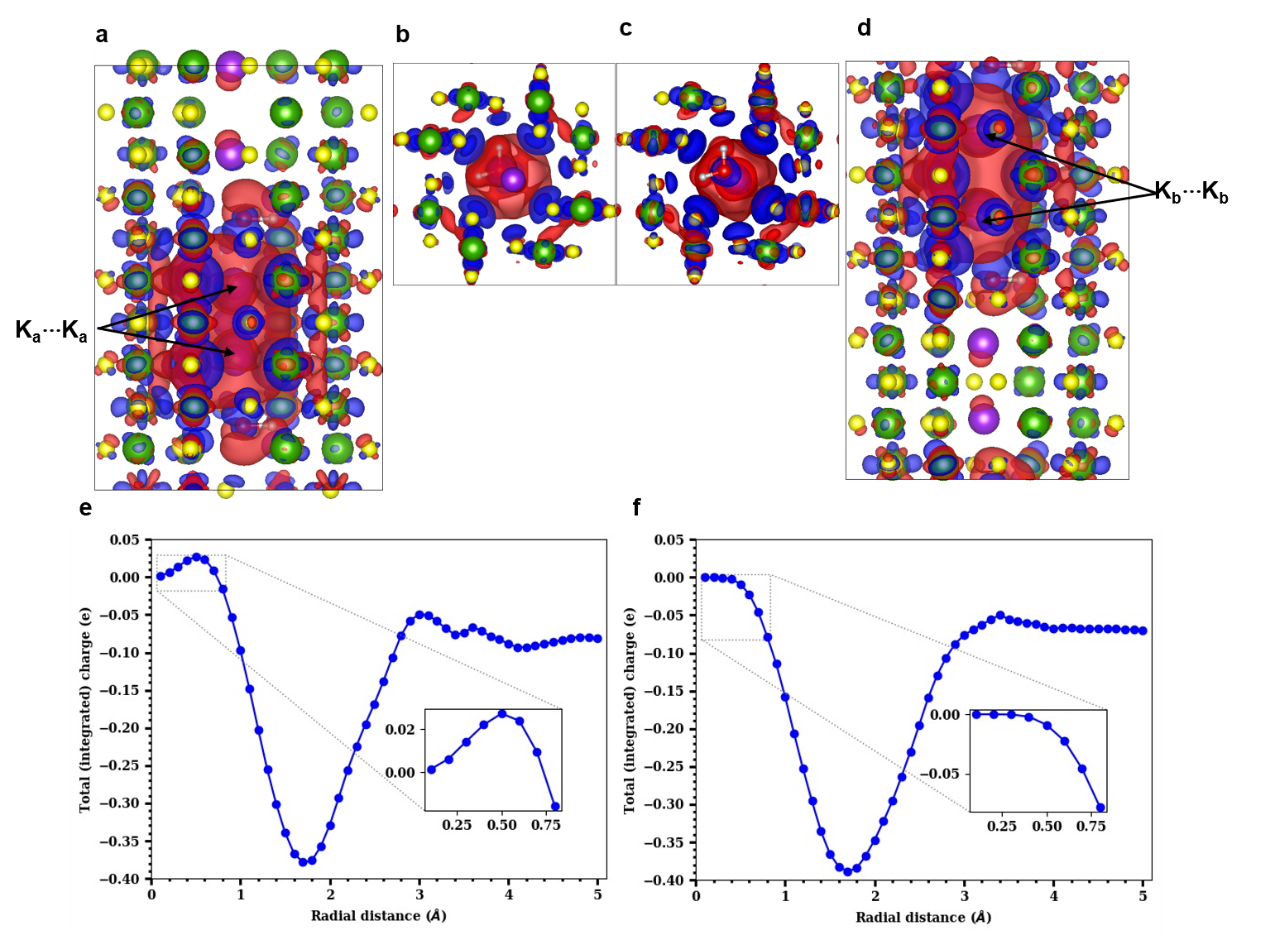}
    \caption{Charge distribution around K$\cdots$K pairs shown by isosurfaces (cutoff = 0.001 e/au$^{-3}$) and integrated charge densities. (a)-(d) Blue regions indicate positive isosurface (electronic charge accumulation); red regions indicate negative isosurface (electronic charge depletion). The charge density of the K$_a$$\cdots$K$_a$ pair occupying ‘2a’ sites is shown along (a) \textit{a}-, and (b) \textit{c}- axes, and of the K$_b$$\cdots$K$_b$ pair occupying ‘2b’ sites is shown along (c) \textit{c}-, and (d) \textit{a}- axes. H$_2$O separates and coordinates with the K$_a$$\cdots$K$_a$ and K$_b$$\cdots$K$_b$ pairs. Integrated charge computed for the (e) K$_a$$\cdots$K$_a$ pair, (f) K$_b$$\cdots$K$_b$ pair.  The insets in (e) and (f) compare the charge behavior in K$_a$$\cdots$K$_a$  and K$_b$$\cdots$K$_b$ pairs.}
    \label{fig:fig5}
\end{figure*}

As already evident in Fig. \ref{fig:fig4}b, some charge is localized between the two K$_a$$\cdots$K$_a$ atoms in the tunnel, but this is not so for the K$_b$$\cdots$K$_b$ pair. To explicate its origin, we analyze the variation in charge densities of the two distinct ion pairs by examining the shared charge along the bonding direction. 
In first-principles electronic structure simulations, charge transfer or charge localization are frequently ill-defined and their accurate determination is nontrivial. This is because the latter requires an arbitrary partitioning or redistribution of the density -- which cannot be done in a unique way -- despite the availability of several approaches for charge localization analyses \cite{ref29}. Hence in what follows, we explicitly compute the difference between electron densities of the fully occupied (K$_{0.80}$Mn$_8$O$_{16}\bm{\cdot}(0.40)$H$_2$O) and the partially filled -- that is, without K$_a$'s (resp. K$_b$'s) -- hollandite tunnels to map the charge distribution due to the K$_a$$\cdots$K$_a$ (resp. K$_b$$\cdots$K$_b$) pair.
Fig. \ref{fig:fig5} shows the charge redistribution when the K$\cdots$K pairs lose their electrons (red isosurface) to the O’s of the hollandite (blue isosurface). The lost electron is shared uniformly by the nearest O’s as depicted by the symmetric blue regions in Fig. \ref{fig:fig5}(a)-(d). The extent of delocalization of the lost charge depends on the K$^+$ position, that is ‘2a’ vs ‘2b’. The ‘2b’ site provides a greater number of nearest O-neighbors and consequently, a higher degree of charge delocalization. The system remains overall charge-neutral. The charge densities around the two H$_2$O’s indicate a strong electrostatic interaction between K$^+$ and H$_2$O. We then compute the integrated charge densities for a region of fixed-length along the $c$-axis, denoted by $Q(r)$, and given by
\begin{align}
    Q(r) &= \int_{\text{H$_2$O}_{\text{below}}+\Delta_1}^{\text{H$_2$O}_{\text{above}}-\Delta_2}dz\int_{x^2 + y^2 \leq r^2}dx\,dy\,\rho(x,y,z) ~,
\end{align}
where the limits $(\text{H$_2$O}_{\text{below}}+\Delta_1, \text{H$_2$O}_{\text{above}}-\Delta_2; \Delta_1 = 0.56 \text{ \AA}, \Delta_2 = 0.68 \text{ \AA})$ exclude the density contributions of the H$_2$O's; $r$ is the radial distance defining the integration cross-section. The two integrated results plotted in Fig. \ref{fig:fig5}e,f, respectively correspond to contributions of the charge density $\rho(x, y, z)$ due to the separate pairs K$_a$$\cdots$K$_a$ and K$_b$$\cdots$K$_b$. A crucial feature of the K$_a$$\cdots$K$_a$ pair is highlighted in the inset of Fig. \ref{fig:fig5}e. In this region, $Q(r)$ takes positive values at short distances ($r < 0.70$ \AA), indicating electronic charge trapped between the two K’s at ‘2a’ sites. By contrast, in Fig. \ref{fig:fig5}f, $Q(r)$ takes negative values at all radial distances -- indicative of electron depletion -- signifying that electronic charge is transferred entirely from the K’s at ‘2b’ sites to the tunneled framework. This comparison of the charge densities for the two K$\cdots$K pairs suggests a mechanism of K$_2^+$ formation when the K’s are confined within the hollandite tunnel at ‘2a’ Wyckoff positions, separated by a small distance ($d_a= 2.62$ \AA), and energetically stabilized by H$_2$O. A part of the electronic charge is then shared between the two K’s resulting in a homonuclear dimer ion. The independent binding energies of the two K’s at ‘2a’ (Fig. \ref{fig:fig5}a) are found to be $-2.988$ eV and $-3.028$ eV. However, the K$_a$$\cdots$K$_a$ pair is more stable with a lower binding energy of $-3.598$ eV. The difference between the average binding energy of the two K’s and the K$_a$$\cdots$K$_a$ pair is used to estimate the K$_a$$\cdots$K$_a$ bond energy, which is E$_{\text{bond}}=0.60$ eV. This supports the formation of the K$_2^+$ -bound species with an internuclear distance of $d_a=2.62$ \AA.

These simulation results confirm and further explain the experimental observations. We note that experiments as well as DFT studies establish the `2b' Wyckoff position as the preferred site for K$^+$ ions inside the hollandite tunnels (see Figs. \ref{fig:fig2}(e), S4 in supplemental material) for low K$^+$ concentrations.  K$^+$ ions are maximally stabilized at the `2b' site due to an optimal coordination from the O's of the surrounding Mn$_8$O$_{16}$ framework. From the energy profile, it is evident that transitioning from a `2b' to a `2a' site requires an uphill climb along the energy surface (see Fig. S6, supplemental material). However, STEM images -- acquired at high dopant concentration -- reveal the displacement of K$^+$ ions to the `2a' sites. Our simulation results demonstrate that this is indeed possible at such high dopant concentration. Two K$^+$ ions are stabilized in the tunnel by the formation of a K$_2^+$ dimer, which arises only when there is enough water in the channel.
Although the bound nature of the ion pair is not accessible to the experiments presented here, the fact that K$^+$ ions occupy the ‘2a’ sites confirms that such a bound pair exists.

\section{Conclusions}
Combining high resolution transmission electron microscopy, and density functional theory calculations, a quasi-one-dimensional water-K$^+$ chain has been identified within one-dimensional tunneled cryptomelane nanorods. We have shown that despite the strong binding energy of K$^+$ ions in well-defined crystallographic sites inside the channel, the presence of water can nevertheless modify coordination sites. They strongly interact with the ions in the channel, altering their transport properties and even reducing the charge of the channels, by facilitating the formation of dicationic molecules which are more difficult to doubly-ionize than single isolated atoms. This is due to both the electrostatic coupling between water molecules and ions, as well as the tunneled structure, which is itself modified in attaining the energy-minima of the system. These results illustrate that the role of water in tunnel-like crystallographic structures of battery materials is significant and should be taken into consideration in such studies. Our results can also provide helpful insights into the mechanisms of emergent functional properties in the field of high-density batteries. Finally, observations about the effects of water stimulate further theoretical discussion about water-related phenomena in tunneled materials.

\begin{acknowledgments}
Electron microscopy work at BNL was supported by the U.S. Department of Energy, Office of Basic Energy Science, Division of Materials Science and Engineering, under Contract No. DE-SC0012704. Material synthesis and the DFT studies associated with the water and potassium occupancy as described in Fig. \ref{fig:fig4} were supported by the Center for Mesoscale Transport Properties, an Energy Frontier Research Center funded by the U.S. Department of Energy, Office of Science, Basic Energy Science, under award $\#$DESC0012673. The DFT studies associated with the K$\cdots$K ion pair distribution as shown in Fig. \ref{fig:fig5} were supported by DOE grant DE-SC0001137. We thank Stony Brook Research Computing and Cyberinfrastructure, and the Institute for Advanced Computational Science at Stony Brook University for access to the high-performance SeaWulf computing system, which was made possible by a \$1.4M National Science Foundation grant (\#1531492).
\end{acknowledgments}

\bibliography{hollandite_refs.bib}
%


\clearpage
\clearpage 
\setcounter{page}{1}
\renewcommand{\thetable}{S\arabic{table}}  
\setcounter{table}{0}
\renewcommand{\thefigure}{S\arabic{figure}}
\setcounter{figure}{0}
\renewcommand{\thesection}{S\arabic{section}}
\setcounter{section}{0}
\renewcommand{\theequation}{S\arabic{equation}}
\setcounter{equation}{0}
\onecolumngrid

\begin{center}
\textbf{Supplemental Material for\\\vspace{0.5 cm}
\large Water-induced formation of an alkali-ion dimer in cryptomelane nanorods\\\vspace{0.3 cm}}
\footnotetext[4]{\label{foot}These authors contributed equally to this work.}

Shaobo Cheng$^{1}$\footnotemark[4], Vidushi Sharma$^{2,3}$\footnotemark[4], Altug S. Poyraz$^{4,5}$, Lijun Wu$^{1}$, Xing Li$^{6}$, Amy C. Marschilok$^{4,6,7}$, Esther S. Takeuchi$^{4,6,7}$, Kenneth J. Takeuchi$^{6,7}$, Marivi Fern\'{a}ndez-Serra$^{2,3}$, and Yimei Zhu$^{1}$.

\small

$^1$\textit{Department of Condensed Matter Physics and Materials Science, Brookhaven National Laboratory, Upton, NY 11973, USA.}

$^2$\textit{Department of Physics and Astronomy, Stony Brook University, Stony Brook, New York 11794-3800, United States}

$^3$\textit{Institute for Advanced Computational Science, Stony Brook University, Stony Brook, New York 11794-5250, United States}

$^4$\textit{Energy Sciences Directorate, Brookhaven National Laboratory, Upton, NY 11973, USA.}

$^5$\textit{Department of Chemistry and Biochemistry, Kennesaw State University, Kennesaw, GA 30144, USA.}

$^6$\textit{Department of Chemistry, Stony Brook University, Stony Brook, NY 11794, USA.}

$^7$\textit{Department of Materials Science and Chemical Engineering, Stony Brook University, Stony Brook, NY 11794, USA.}

(Dated: \today)
\end{center}

\begin{figure}[H]
    \centering
    \includegraphics[width=3.in]{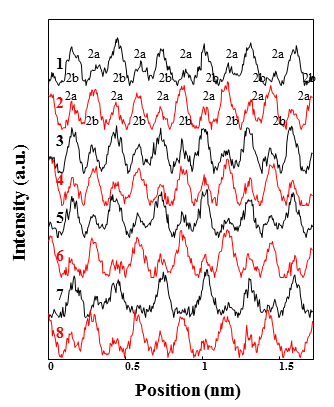}
    \caption{Intensity profiles from eight vertical tunnels in Fig. 2b, showing intensity variation at `2a' sites. Note that the background has been subtracted for comparison.}
    \label{fig:figS1}
\end{figure}

\begin{figure}[H]
    \centering
    \includegraphics[width=4in]{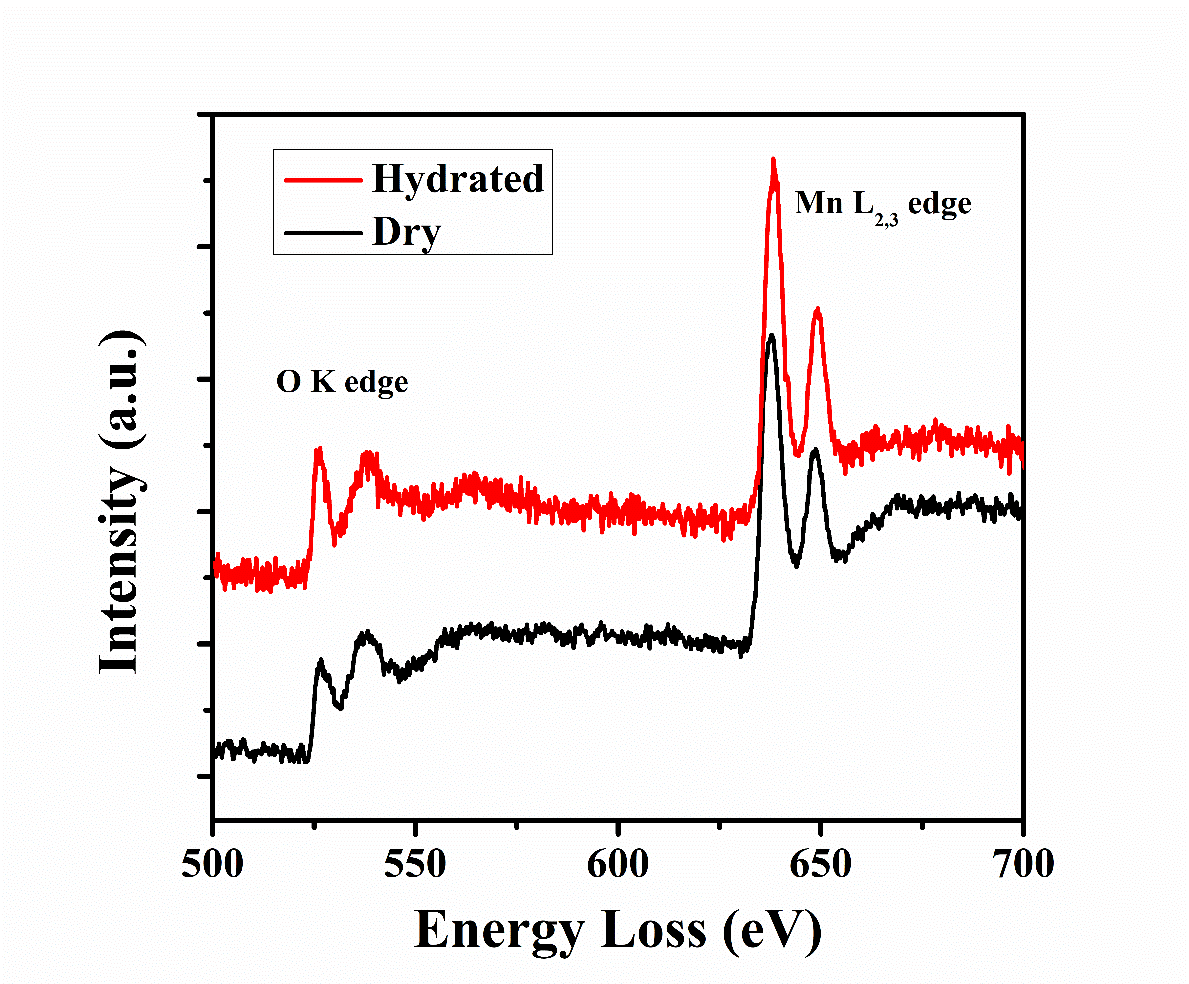}
    \caption{Core loss electron energy-loss spectroscopy showing the dry and hydrated samples. The prepeak of O K edge in manganites reflects the concentration of oxygen in the materials. The prepeak from the hydrated sample is higher than that from the dry sample, indicating the presence of more water in the hydrated sample.}
    \label{fig:figS2_new}
\end{figure}

\begin{figure}[H]
    \centering
    \includegraphics[width=3.5in]{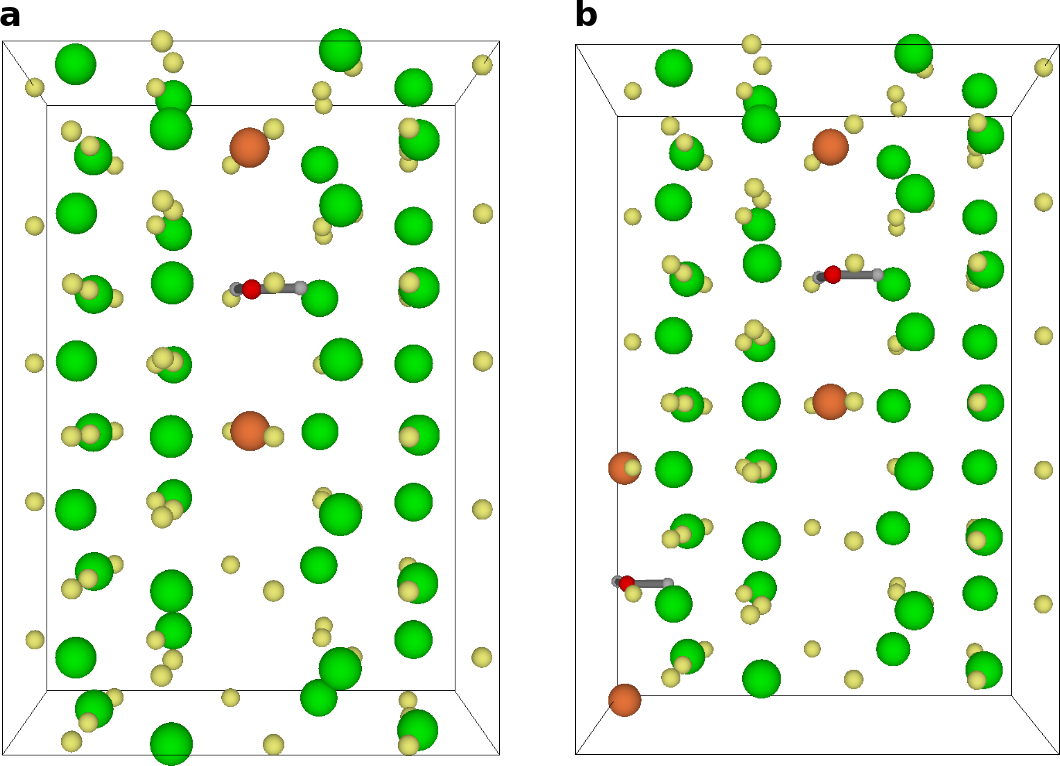}
    \caption{\ce{K_{0.40}Mn8O16 * {(0.20)} H2O}: Two \ce{K+}'s and one \ce{H2O} occupying (a) \textbf{one} tunnel and, (b) \textbf{both} the tunnels of five unit cells of $\alpha$-\ce{Mn8O16}, shown along the $a-$axis. All atomic species occupy ‘2b’ sites inside the tunnel(s). The binding energy of \ce{K+} is (a) $-4.723$ eV and (b) $-4.706$ eV.}
    \label{fig:figS8}
\end{figure}

\begin{figure}[H]
    \centering
    \includegraphics[width=4.5in]{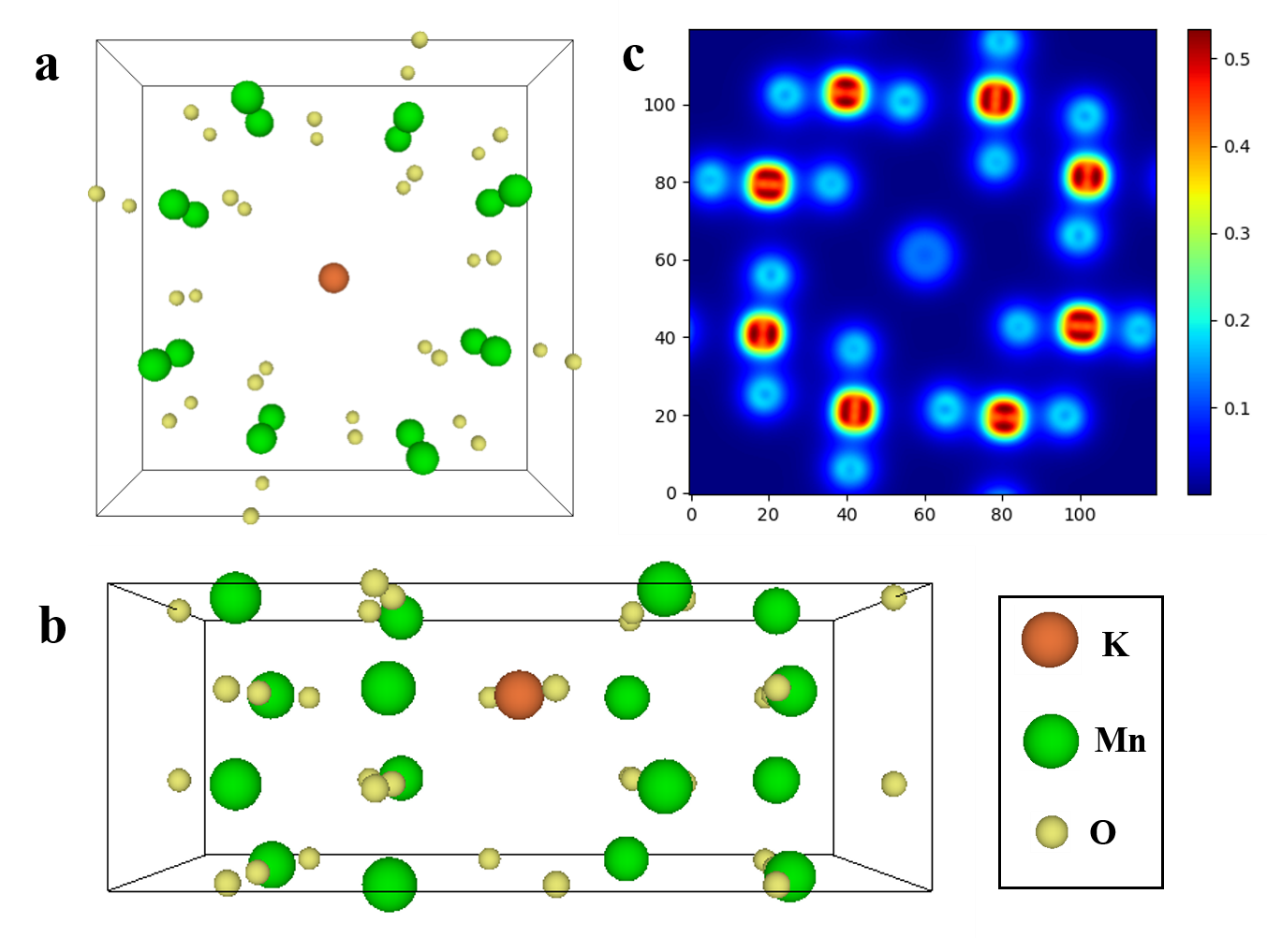}
    \caption{\ce{K_{0.50}Mn8O16}: K-doped two unit cells of $\alpha$-\ce{Mn8O16}, shown along (a) the c-axis, and (b) the a-axis. \ce{K+} attains a ‘2b’ Wyckoff position inside the $[2\times 2]$ hollandite tunnel, surrounded by eight O’s lying in the same plane as \ce{K+} itself. (c) Mean charge density plot of K-doped hollandite from the $[001]$ axis showing \ce{K+} occupying the tunnel centered site. The binding energy of \ce{K+} (BE$_\text{K}$) is computed to be $-4.528$ eV.}
    \label{fig:figS2}
\end{figure}

\begin{figure}[H]
    \centering
    \includegraphics[width=5.0in]{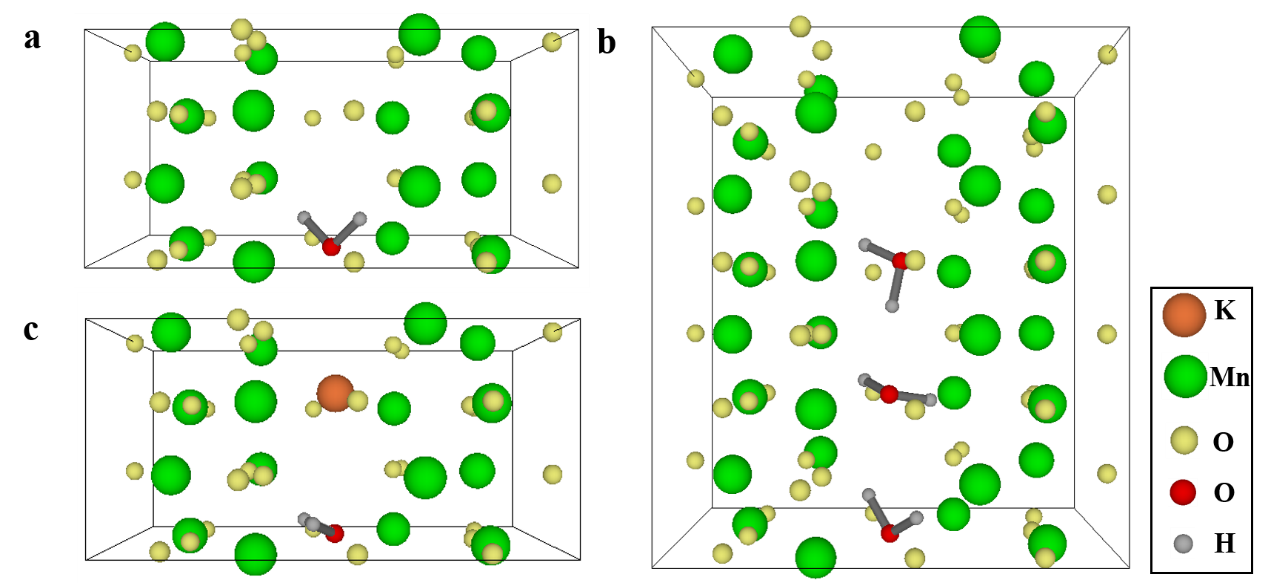}
    \caption{DFT calculated results for hollandite cells containing water. (a) \ce{Mn8O16 * {(0.50)} H2O}. \ce{H2O} in two unit cells of $\alpha$-\ce{Mn8O16}, shown along the a-axis. \ce{H2O} attains a ‘2b’ Wyckoff position inside the $[2\times 2]$ hollandite tunnel. The binding energy of \ce{H2O} is computed to be $-0.212$ eV. (b) \ce{Mn8O16 * {(0.75)} H2O}. Three molecules of \ce{H2O} in four unit cells of $\alpha$-\ce{Mn8O16}, shown along the a-axis. H2O’s attain ‘2b’ Wyckoff positions inside the  hollandite tunnel. The binding energy of \ce{H2O} is computed to be $-0.475$ eV. H2O’s form a H-bond chain spiraling down the axis of the tunnel (also the c-axis). (c) \ce{K_{0.50}Mn8O16 * {(0.50)} H2O}. \ce{K+} and \ce{H2O} in two unit cells of $\alpha$-\ce{Mn8O16}, shown along the a-axis. \ce{K+} and \ce{H2O} occupy ‘2b’ Wyckoff positions inside the $[2\times 2]$  hollandite tunnel. The binding energy of K is $-5.096 eV$; binding energy of \ce{H2O} is $-0.772$ eV.}
    \label{fig:figS3}
\end{figure}

\begin{figure}[H]
    \centering
    \includegraphics[width=4.0in]{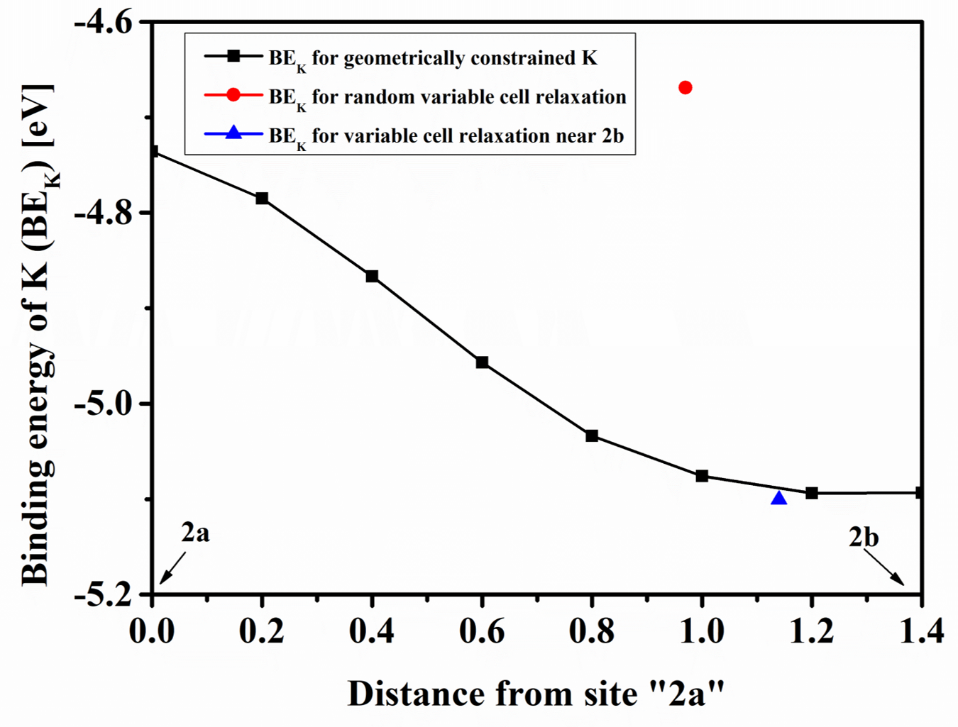}
    \caption{Binding energy of \ce{K+} for constrained positions (from ‘2a’ to ‘2b’) of K+ inside the $[2\times 2]$ tunnel of \ce{K_{0.50}Mn8O16 * {(0.5)} H2O}. The black curve indicates the variation in the binding energy of  \ce{K+} from a maximum at ‘2a’ to a minimum at ‘2b’, while water in the tunnel sits at ‘2b’. The blue point denotes the position attained by \ce{K+} under a variable cell-relaxation of the structure when the initial position of \ce{K+} is near ‘2b’. The red point represents a final position of \ce{K+} after a variable cell-relaxation when it was initially randomly placed, indicating the presence of several energy troughs along the tunnel axis.}
    \label{fig:figS4}
\end{figure}

\begin{figure}[H]
    \centering
    \includegraphics[width=2.5in]{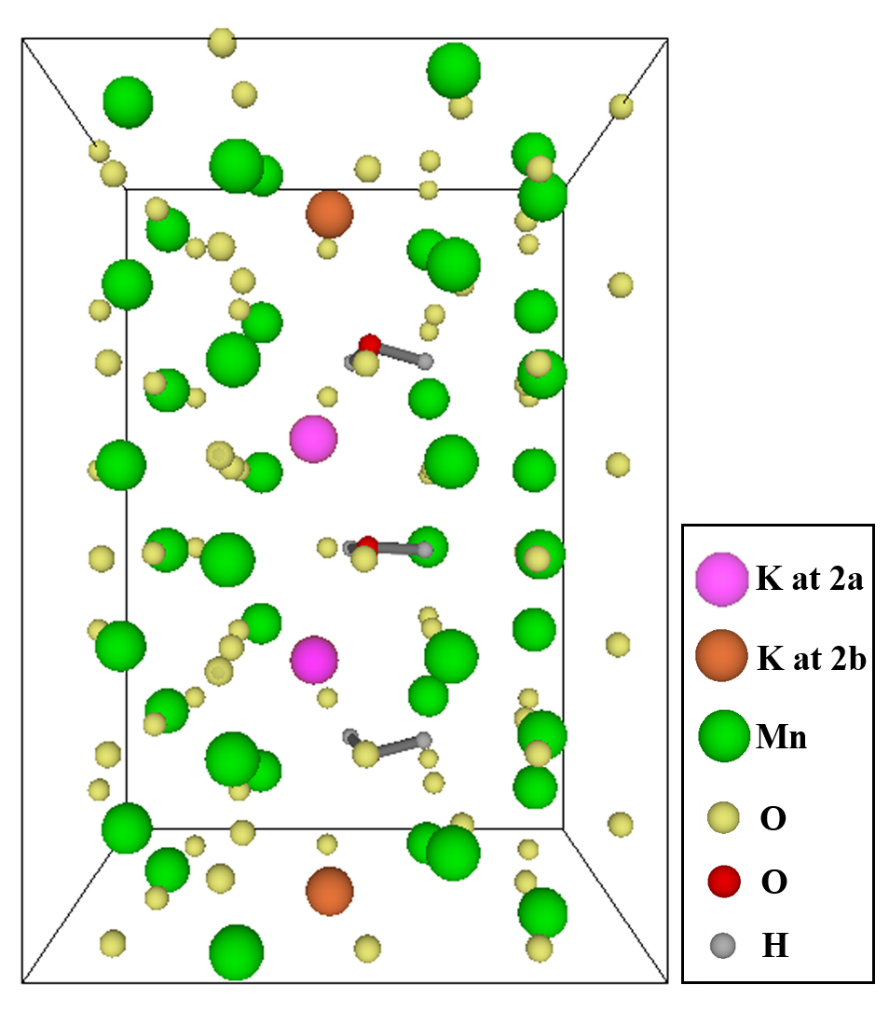}
    \caption{\ce{K_{0.80}Mn8O16 * {(0.60)} H2O}: Four \ce{K+} and three \ce{H2O} in five unit cells of $\alpha$-\ce{Mn8O16}, shown along the a-axis. The two outer \ce{K+} ions occupy ‘2b’ sites, and the two inner \ce{K+} ions are displaced by \ce{H2O}’s to occupy positions near ‘2a’. The \ce{H2O} trapped between \ce{K+} solvates it. The binding energy of \ce{K+} is $-2.996$ eV. The higher binding energy accounts for the cost paid for overpopulating the tunnel.}
    \label{fig:figS5}
\end{figure}

\newpage

\begin{figure}[H]
    \centering
    \includegraphics[width=6.5in]{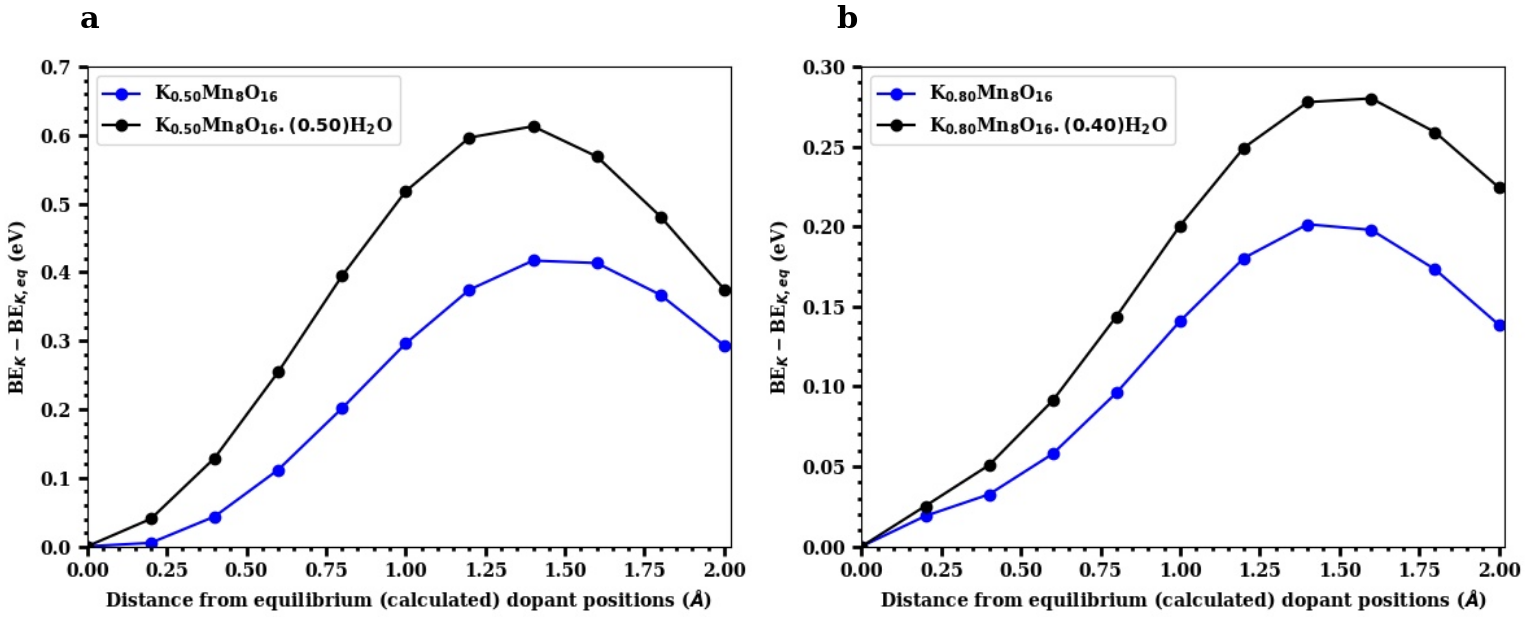}
    \caption{Energetics of translating the dopants along the \textit{c}-axis of the $\alpha$-\ce{Mn8O16} tunnel. The starting point is the equilibrium position of (all) dopants in the tunnel as obtained by our relaxation studies on the structures. The dopants are then uniformly shifted along the axis in intervals of 0.2 \AA\, and the binding energy of potassium relative to that in the equilibrium position (BE$_{\text{K}}$-BE$_{\text{K,eq}}$ [eV]) is computed for (a) low concentration of dopants with (\ce{K_{0.50}Mn8O16 * {(0.50)} H2O} and without water (\ce{K_{0.50}Mn8O16}), and (b) experimentally relevant - high concentration of dopants with (\ce{K_{0.80}Mn8O16 * {(0.40)} H2O}) and without water (\ce{K_{0.80}Mn8O16}).}
    \label{fig:figS6}
\end{figure}

To illustrate the effect of \ce{H2O} on the movement of K along the tunnel, we estimate the energy barrier encountered by K both in the presence of water (black curve) and in its absence (blue curve) for two different concentrations of dopants as shown in Fig. \ref{fig:figS6}. The computed energy barriers are an upper bound to the actual barriers, because we do not conduct explicit barrier calculations. Instead we allow the entire system to relax, constraining the dopant position in the channel. In the low-dopant concentration regime Fig. \ref{fig:figS6}(a), the starting equilibrium configuration of dopants has K and \ce{H2O} situated at separate `2b' sites and as we continue to move these dopants from their respective favorable positions, the binding energy of K continues to increase reaching its maximum where K and \ce{H2O} are both located at different `2a' positions, yielding a barrier of $\approx$0.42 eV for the case with no water and $\approx$0.60 eV for the structure containing water. This seems to further suggest that the presence of water mostly stabilizes \ce{K+} at its original `2b' position, making its departure from the equilibrium position less favorable. Furthermore, in the experimentally-relevant high concentration limit, Fig. \ref{fig:figS6}(b), the initial equilibrium dopant configuration now has K’s occupying a mix of `2b' and `2a' sites and translating all the dopants in the tunnel now results in a binding energy maximum where the K’s effectively exchange their `2b' and `2a' positions, with the inner K’s now at `2b' and the outer ones at `2a'. The energy barrier for K in this densely occupied tunnel is $\approx$0.20 eV in the absence of water and $\approx$0.28 eV for water present in the tunnel. This is further evidence that though the presence of excess K’s  improves the migration through the tunnel, \ce{H2O} still continues to stabilize the K’s by not lowering the energy barrier. Thus, \ce{H2O} does not facilitate the transport of K’s inside the $\alpha$-\ce{Mn8O16} tunnels, and tends to overall stabilize K’s in the originally chosen favorable coordination environment by forming a solvation shell around them.

\end{document}